\documentclass[journal]{vgtc}                     

\onlineid{1146}

\vgtccategory{Research}

\vgtcpapertype{application/design study}

\title{Amplifying the Music Listening Experience through Song Comments on Music Streaming Platforms}

\author{%
  Longfei Chen, Qianyu Liu, Chenyang Zhang, Yangkun Huang, \\ \authororcid{Zhenhui Peng}{0000-0002-5700-3136}, \authororcid{Haipeng Zeng}{0000-0002-0339-0361}, \authororcid{Zhida Sun}{0000-0003-4689-986X}, \authororcid{Xiaojuan Ma}{0000-0002-9847-7784}, and \authororcid{Quan Li}{0000-0003-2249-0728}
}

\authorfooter{
  \item
  	Longfei Chen, Qianyu Liu, and Quan Li are with the School of Information Science and Technology, ShanghaiTech University, and Shanghai Engineering Research Center of Intelligent Vision and Imaging, China. Quan Li is the corresponding author. E-mail: \{chenlf,liuqy2022,liquan\}@shanghaitech.edu.cn.
  \item
  	Chenyang Zhang is with the Department of Computer Science, University of Illinois at Urbana-Champaign. E-mail: zhang414@illinois.edu.
   \item Yangkun Huang is with Tandon School of Engineering, New York University. E-mail: yh1875@nyu.edu.
  \item Zhenhui Peng is with the School of Artificial Intelligence, Sun Yat-Sen University. E-mail: pengzhh29@mail.sysu.edu.cn.
  \item Haipeng Zeng is with the School of Intelligent Systems Engineering, Sun Yat-Sen University. E-mail: zenghp5@mail.sysu.edu.cn.
  \item Zhida Sun is with the College of Computer Science and Software Engineering, Shenzhen University. E-mail: zhida.sun@outlook.com.
  \item Xiaojuan Ma is with the Department of Computer Science and Engineering, the Hong Kong University of Science and Technology. E-mail: mxj@cse.ust.hk.
}

\abstract{%
Music streaming services are increasingly popular among younger generations who seek social experiences through personal expression and sharing of subjective feelings in comments. However, such emotional aspects are often ignored by current platforms, which affects the listeners' ability to find music that triggers specific personal feelings. To address this gap, this study proposes a novel approach that leverages deep learning methods to capture contextual keywords, sentiments, and induced mechanisms from song comments. The study augments a current music app with two features, including the presentation of tags that best represent song comments and a novel map metaphor that reorganizes song comments based on chronological order, content, and sentiment. The effectiveness of the proposed approach is validated through a usage scenario and a user study that demonstrate its capability to improve the user experience of exploring songs and browsing comments of interest. This study contributes to the advancement of music streaming services by providing a more personalized and emotionally rich music experience for younger generations.
}

\keywords{Music comments, music streaming services, visualization}

\graphicspath{{figs/}{figures/}{pictures/}{images/}{./}} 

\usepackage{tabu}                      
\usepackage{booktabs}                  
\usepackage{lipsum}                    
\usepackage{mwe}                       
\usepackage{color}
\usepackage{CJKutf8}
\usepackage{multirow}
\usepackage{graphicx}
\usepackage{balance}
\usepackage{paralist}
\usepackage{array}
\usepackage{booktabs}
\usepackage{wrapfig}
\usepackage{tabularx}
\usepackage{mathptmx}                  

\newcommand{\clf}{\textcolor{black}}

\begin{document}


\firstsection{Introduction}

\maketitle

\par The integration of music into individuals' daily lives has become increasingly prevalent. In particular, the emergence of music streaming platforms, including \textit{YouTube Music} and \textit{NetEase Cloud Music}, has provided users with not only fundamental music listening capabilities but also a forum-like comment section that enables listeners to share their sentiments, preferences, and personal narratives. These social functionalities have significantly altered the mode of music consumption and have facilitated connections between individuals with similar musical preferences, overcoming geographical constraints~\cite{wang:2020:exploring,chen:2018:research}.


\par The act of leaving comments on music has become a widespread means for listeners to express their agreement, sympathy, or disagreement with others by sharing their experiences in embedded forums on music streaming platforms or by responding or ``liking'' other users' comments. In this study, we extract listeners' interests, preferences, and emotions toward songs, as well as related personal experiences from a vast number of music comments. The results obtained have the potential to assist users in exploring songs with specific emotions (e.g., happiness or homesickness) or in particular scenarios (e.g., work and parties). In such instances, users can peruse a set of song lists labeled with various keywords provided by the platform or other users that correspond to a specific mood or situation. Subsequently, the user may select a specific song from the list to listen to, either randomly or based on prior experience. While listening to the song, the user can also browse comments from other listeners, some of which may be of interest to them. We refer to this process as the ``Library $\rightarrow$ Song List $\rightarrow$ Song $\rightarrow$ Comment'' exploration process. However, exploring music comments to help users find songs and comments of interest is not a trivial task for several reasons. First, existing song lists on most platforms only show basic music metadata, such as the song's title, singer, and album name, while comments, which are the primary means of conveying personal feelings, are not readily accessible to users in the song list interface. As a result, it is challenging for users to identify songs in a given list that may elicit the appropriate nuances of their emotions. Second, the manner in which comment sections are organized on current music streaming platforms makes it difficult for listeners to locate posts that are relevant to their interests. Most platforms only offer two types of comment filtering: ``sort by time'' and ``sort by popularity.'' ``Sort by time'' tends to neglect early comments, while ``sort by popularity'' tends to overlook minority comments. A considerable number of comments are arranged in chronological order or sorted by popularity, requiring users to sift through them one by one to find the ones that interest them. Third, music comments can be complex and convoluted. Comments over time can be intertwined to address several topics, some of which are not entirely related to the song itself. For instance, ardent fans may boast about their group identity, which may be unrelated to the music~\cite{ren:2012:building,sugiana:2018:construction}. Specific events, such as Michael Jackson's passing, may also spark discussions in comments~\cite{siddiqi2015keyword}.

\par In order to assess the viability of the concept of enhancing songs with information derived from music comments, aimed at aiding users in the discovery of songs and comments that pique their interest, we carried out a questionnaire involving a cohort of 104 music streaming service users who were actively engaged. The results revealed that 75.49\% of the participants expressed a keen interest in the integration of a music comment feature within existing platforms, with 61.04\% of them having actively contributed comments themselves. Moreover, an additional 58.82\% of the respondents held the belief that prevalent mainstream platforms inadequately acknowledge the significance of comments in enhancing the overall user experience. In this study, we introduce two features that enhance an existing music streaming application. These features are the comment-related preview tags for songs and map metaphor visualization of comments. In order to generate the preview tags for each song in the song list, language models (LM) are used to analyze all the comments. This analysis includes the extraction of high-frequency keywords, topic detection, sentiment analysis, and identification of the induced mechanism. The eight preview tags for each song are then displayed on the song list page. The aim of this feature is to optimize user experience in the ``Song List $\longrightarrow$ Song'' subprocess and provide a personalized music exploration for song seekers. The comment map feature on the comment details page structures and summarizes the comments, which facilitates the identification of personally relevant interests in the ``Song $\rightarrow$ Comment'' subprocess. The proposed approach undergoes evaluation via a usage scenario, followed by a user study assessing the usefulness, effectiveness, user interactions, and impact of our approach. The primary contributions can be summarized as follows.
\begin{compactitem}
\item We conducted a formative study aimed at comprehending the requirements of users when utilizing music streaming platforms and capturing their browsing experiences with music comments.
    \item We fine-tuned pre-trained LMs to extract keywords of high frequency, identify topics, assess sentiment, and determine induced mechanisms from chronologically organized music comments. Subsequently, we enhanced an existing music streaming platform with innovative visual designs to empower end-users with a comprehensive overview of song comments and their dynamics.
    \item We conducted a usage scenario and a user study involving end-users to validate the efficacy of our proposed approach. Through these activities, we derived valuable insights and presented evidence demonstrating the effectiveness of our approach.
\end{compactitem}

\section{Related Work}

\subsection{Sentiments and Induced Mechanisms Behind Music Comments}
\par Over the past decades, there has been significant attention given to empirical studies on musical sentiments~\cite{marin2010music,zeng2020emotioncues,zeng2019emoco}. These studies have consistently reported that music can effectively communicate sentiments to the listener and may also induce the listener's own sentiments. For example, Marin et al.~\cite{marin2010music} suggest that ``\textit{automatic affective responses to stimuli are essentially relevant for subsequent cognitive, emotional, and behavioral reactions: emotions induced by music crucially influence the (emotional) processing of pictures, facial expressions, films, and words.}'' As a result, listeners' comments on music reflect the sentiments they experience towards the song, which may be a result of the music itself or the induced effects of the music.

\par To comprehend what triggers musical sentiments, the field of psychology has developed two complex explanatory studies based on human cognitive theory. The first study by Schere et al.~\cite{scherer2001emotional} proposes a broad interaction profile based on the Component Process Model (CPM), which assumes that there are three emotions associated with music and that experts can calculate the contribution of four factors (structural, performance, listener, and contextual features) in inducing sentiments. However, the weights of the parameters involved in the above factors require experimental data to be calculated and are challenging to quantify directly from the comment text. Alternatively, Juslin et al.~\cite{JUSLIN:2013:everyday} proposed the BRECVEMA framework, which summarizes eight induced mechanisms, at least one of which generates musical sentiments. In this work, we incorporate the underlying sentiment and the induced mechanism as additional properties of the comment text to help users better explore the comments of a song.

\subsection{Comment Data Processing}

\par The majority of studies on comment data processing have been focused on sentiment analysis~\cite{pang2002thumbs}. However, binary classification of positive or negative sentiments is insufficient for providing cues for further human understanding. Following the conventional sentiment classification by Chuang et al.\cite{chuang2004multi,eckman1972universal}, we categorize sentiments into six types: \textit{angry}, \textit{neutral}, \textit{sad}, \textit{fear}, \textit{surprise}, and \textit{happy}. We then utilize fine-tuned deep-learning models to predict the sentiment category of each comment. Similarly, we employ another fine-tuned model to predict the category of the induced mechanism. \clf{In order to categorize the induced mechanisms, we make use of the \textit{BRECVEMA} framework~\cite{JUSLIN:2013:everyday}, originally introduced by Juslin. Our approach involves referencing and aligning with this framework for efficient classification.} 

For the specific scenario of music comment analysis, we simplify the eight mechanisms proposed in~\cite{JUSLIN:2013:everyday} into four categories: 1) \textit{Music Evaluation.} The lyrics, melody, rhythm, performance style, recording mode, and production team (excluding the singer) influence the quality of the music clip. Listeners express their immediate attitudes, evoke visual imagery, and make aesthetic judgments about a song through music comments that refer to the above aspects. 2) \textit{Personal Memories and Experiences.} A song may trigger a listener's personal memories of an event in their life. 3) \textit{Contextual Information.} Comments on a song made by listeners are not only related to the song itself, but they also refer to corresponding social events and human knowledge that may be relevant to a wider range of people. 4) \textit{Others.} Some music comments are incomprehensible and philosophical, while others involve listeners copying and pasting lyrics they enjoy. These comments cannot be classified into the aforementioned three categories.

\par Moreover, comment analysis research has also investigated methods for detecting keywords~\cite{bharti2017automatic,hasan2014automatic}. Statistical-based techniques utilize frequency measures to choose the top $n$ candidates based on the linguistic corpus. Graph-based methods~\cite{beliga2015overview} use bag-of-words (BOW) with co-occurrence metrics, which generates an $n$-dimensional vector for each document. Linguistic approaches utilize the linguistic features of the words for keyword detection~\cite{barzilay1999using,hulth2003improved}. \clf{Nevertheless, the aforementioned techniques are constrained by their reliance on surface-level word frequency, statistics, or rule-based approaches, limiting their capacity to capture intricate semantic information. Notably, various machine learning methods, including Na\"ive Bayes~\cite{uzun2005keyword}, SVM~\cite{cortes1995support}, HMM~\cite{baum1970maximization}, and CRF~\cite{lafferty2001conditional}, have been explored for keyword extraction. However, these methodologies often treat words as independent features and struggle to capture the intricate relationships and contextual dependencies between them. In contrast, deep learning methods~\cite{graves2013hybrid,graves2005framewise,zhang2016keyphrase} exhibit enhanced capability in extracting keywords by effectively harnessing contextual semantic information. These models can automatically learn linguistic patterns and contextual relationships within the text, thereby offering a more comprehensive understanding of the underlying data. By incorporating attention mechanisms, contextual encoding, and other techniques, these deep learning models further demonstrate improved accuracy in inferring and predicting the significance of individual words. To annotate keywords based on their contextual relevance to the given sentence, our study employs a self-supervised deep learning approach~\cite{grootendorst2020keybert}.}



\subsection{Visualization of Sentiments and Topics in Text}
\par Over the past decade, a diverse range of visualization techniques has emerged for the analysis of sentiment and thematic patterns in text, encompassing both simplistic infographics and sophisticated visual analytics systems~\cite{kucher2018state}. For example, one such approach known as \textit{Review Spotlight}\cite{yatani2011review} utilizes a tag cloud consisting of prominent adjective/noun pairs to provide a summary of customer reviews. Another technique, referred to as \textit{OpinionBlocks}~\cite{alper2011opinionblocks}, employs multiple coordination bars and text tags to explore polarities and noteworthy keywords associated with specific product features. \clf{However, these existing methods fail to adequately capture the temporal evolution of comments.} In addition, Zhao et al.~\cite{zhao2014pearl} introduced a technique called \textit{PEARL} to analyze the sentiment changes of individual users on social media over time. \clf{However, this approach falls short in the context of our study, as the analysis of a single user is insufficient to address our requirements.}


\begin{table*}[h]
\center
\caption{In the questionnaire, a total of $102$ valid responses were obtained. The participants were divided into two groups based on their commenting habits. Participants on the left were found to be accustomed to browsing comments, and the distribution of the 5-point Likert scale question on the $13$ comment contents, as well as the other two usage habits, were presented. On the other hand, participants on the right were found to rarely use the commenting function, and their possible motivations were shown.}
\label{tab:survey}
\begin{tabular}{cccccccc}
\hline
\multicolumn{8}{c}{\textbf{Have the habit of following and using commenting function}}   \\
\multicolumn{6}{c}{\textbf{Yes}}                                                         & \multicolumn{2}{c}{\textbf{No}}                                                     \\
\multicolumn{6}{c}{77/102}                                                                                                                                                                        & \multicolumn{2}{c}{25/102}                                                          \\ \hline
\textbf{comment type} & \textbf{AVG}             & \textbf{SD}                                   & \textbf{comment type} & \textbf{AVG}     & \multicolumn{1}{c|}{\textbf{SD}}      & \textbf{}                                               &                           \\ \cline{1-6}
Background                    & 4.25                     & 0.93                                          & Latest                      & 2.56             & \multicolumn{1}{c|}{1.03}             & \multicolumn{2}{c}{\textbf{Dislike comment because}}                                \\
Professional evaluation       & 4.16                     & 0.97                                          & Personal story              & 2.49             & \multicolumn{1}{c|}{1.23}             & \multicolumn{1}{r}{}                                    &                           \\
Similar   Feeling             & 3.92                     & 0.93                                          & Live content                & 2.27             & \multicolumn{1}{c|}{1.14}             & \multicolumn{1}{l}{Massive data with no classification}                & \multicolumn{1}{r}{18/25} \\
Most   popular                & 3.56                     & 0.97                                          & Social news                 & 2.08             & \multicolumn{1}{c|}{1.09}             & \multicolumn{1}{l}{No content of interest}              & \multicolumn{1}{r}{10/25} \\
Info of   creative team       & 3.47                     & 1.21                                          & (Anti-)fans                 & 1.90             & \multicolumn{1}{c|}{0.97}             & \multicolumn{1}{l}{Affect experience of listening}       & \multicolumn{1}{r}{6/25}  \\
Literature   creation         & 2.90                     & 1.28                                          & Terse comment               & 1.57             & \multicolumn{1}{c|}{0.85}             & \multicolumn{1}{l}{No habit of browsing} & \multicolumn{1}{r}{5/25}  \\
Fancy   content               & 2.86                     & 1.23                                          & \textbackslash{}            & \textbackslash{} & \multicolumn{1}{c|}{\textbackslash{}} & \multicolumn{1}{l}{Unable to empathize}                 & \multicolumn{1}{r}{5/25}  \\ \cline{1-6} \hline

\end{tabular}
\end{table*}

\par The visualization of topics has been a widely explored area in the literature, typically using stacked graphs~\cite{byron2008stacked,dork2010visual,havre2002themeriver,leskovec2009meme}. Interactive visualizations have been proposed by \textit{TIARA}~\cite{liu2009interactive,wei2010tiara}, which combines text summarization with interactive visualization, and \textit{Textflow}~\cite{cui2011textflow}, which employs a river-based visualization to display merging and splitting relationships between topics. Word clouds with sentiment analysis have also been utilized for comment visualization~\cite{kucher2018state,wang2018towards}. Chen et al.~\cite{Chen:2016:D-Map,Chen:2017:E-Map,chen2019r} developed a structured semantic space for exploring the social media model of ego-centric information diffusion and event evolution. In this approach, users involved in retweeting central posts are mapped to a hexagonal grid based on the similarity and temporal order of their behavior, with the edges of each hexagonal cell constructed to follow the retweets and replies between them. However, in the case of music comments, most of them are original with no forwarding relationships, making it challenging to construct relationships and track the evolution. Motivated by the aforementioned studies, we propose a novel visualization approach that utilizes the map metaphor to restructure music comments.

\section{Formative Study}
\par In order to better understand the actual needs of users when using music streaming platforms and the process by which users connect through music comments, we conducted a formative study, wherein participants were invited to respond to a questionnaire that aimed to elicit their experiences in browsing music comments.

\subsection{Procedure}
\par We designed a questionnaire to discover users' usage scenarios and preferences when viewing, receiving, analyzing, and writing music comments. First, we designed questions to gather information on users' usage scenarios and preferences regarding viewing, receiving, analyzing, and writing music comments. Initially, the questionnaire probed participants on their usage habits of the commenting function and preferences for various types of music comments. The comment data was subdivided into $13$ sub-items, each of which was assessed using a Likert scale. To determine the positivity of browsing music comments, participants rated a variety of comment types on a scale of $1$ to $5$, ranging from ``don't want to see it at all'' to ``very much want to see it''. The questionnaire also investigated the interaction between commenting and listening to songs. Additionally, we identified the information carried by songs on multiple music platforms and asked users to rank them from highest to lowest based on their level of interest. Finally, the questionnaire investigated the user interface design that would impact the user's listening experience.

\subsection{Participants}

\par Our study recruited a total of $104$ participants through social media posts and received 102 responses. Of these respondents, $64$ were male, $37$ were female, and one preferred not to specify their gender. The majority of respondents fell within the $16-28$ age group, with $78$ participants, while $21$ participants fell within the $29-45$ age group, and only two were above the age of $45$. Educational levels varied, with one participant having a junior high school education, three having a high school education, $65$ having an undergraduate degree, $29$ having a postgraduate degree, and four holding a Ph.D. The demographic distribution of our sample suggests that our results are valid. Specifically, younger and middle-aged individuals tend to use music streaming services more frequently and seek positive social experiences. To further investigate listeners' challenges and requirements in using music streaming services, we conducted semi-structured interviews with four participants, referred to as P1 to P4. These participants were selected based on their frequent use of music streaming services and varied commenting habits. P1, P2, and P3 were music enthusiasts who were knowledgeable about at least one genre of music and ranged in age from $20$ to $23$. P4 was an amateur songwriter registered on NetEase Cloud Music and had released about $10$ songs in three years.

\par We conducted the interviews using a teleconferencing tool and audio-recorded them with consent. The interviews lasted between $30$ and $50$ minutes and were divided into three parts, similar to the questionnaire. We followed an iterative coding process~\cite{hruschka:2004:reliability} to analyze the usage habits and needs of streaming music users. Two researchers coded the four interview transcripts separately, using a codebook for text-based questions. We then calculated Cohen's kappa to assess inter-rater reliability, with the average score across all codes being $0.87$ (SD=$0.10$, ranging from $0.74$ to $1.00$) and an average of $93.7\%$ agreement. After independent coding, the researchers discussed interpretations and discrepancies until they reached a consensus on the codebook, then adjusted their code data.

\begin{figure*}[h]
 \centering 
 \includegraphics[width=\linewidth]{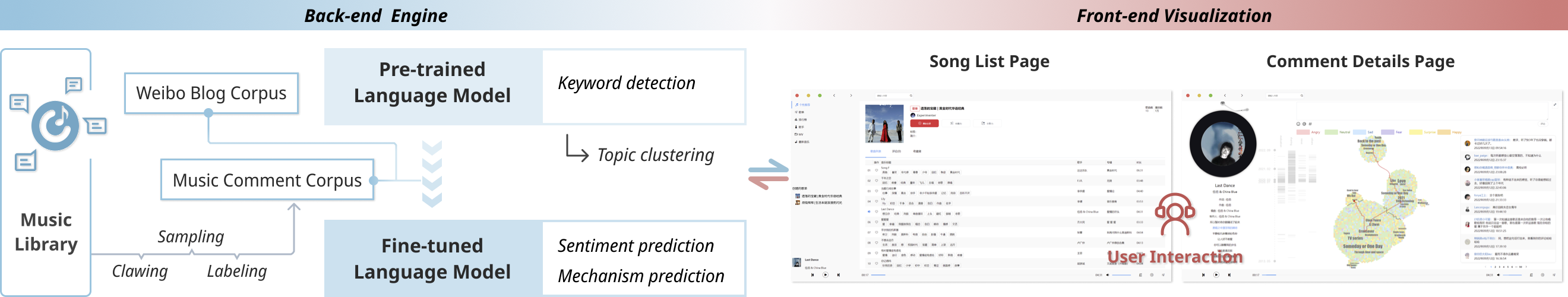}
 \caption{The pipeline of the architecture consists of a data processing module, a back-end engine, and a front-end visualization interface.}
 \label{fig:pipeline}
\end{figure*}

\subsection{Results}

\par The questionnaire findings are presented in \autoref{tab:survey}. The results revealed a number of challenges that the participants encountered while utilizing the commenting feature on music streaming platforms.

\par \noindent\textbf{C.1 Users' interests are variable and need-dependent.} \autoref{tab:survey} depicts the general preference of individuals regarding comments. More than half of the participants showed a preference for neutral content, such as expertise and music appreciation. Additionally, 75\% of this group expressed a lack of interest in comments that convey emotions. However, this trend is not absolute, particularly when specific scenarios are simulated, such as loss of love or separation, resulting in significant differences in participants' choices. Furthermore, P4 noted a similar situation, stating that ``\textit{as a music creator, I would prefer to see comments that point out problems with my songs than unjustified praise. But when I am blocked from creating and depressed, negative comments tend to make me self-doubt and I would feel that reading some positive comments would be a better choice.}''


\par \noindent\textbf{C.2 Massive comments pile up, making it difficult for users to get to the content of interest.} Approximately 25\% of the questionnaire participants exhibit low levels of interest in music comments, while nearly 66.7\% avoid perusing comments altogether due to negative experiences stemming from the current form of comment organization. Specifically, the arduous task of finding relevant comments of interest, coupled with the high prevalence of extraneous and superfluous information, which includes fan comments, represents a significant obstacle to the enjoyment of music. In agreement with these sentiments, P1 articulated that the current comment section can be a source of frustration, stating that ``\textit{everyone says whatever they want on the Internet,}'' leading to futile discussions.

\par \noindent\textbf{C.3 Existing platforms ignore the impact of comments on users' decisions.} According to the findings, users who follow comments typically engage with them while listening to music. Additionally, due to the product interaction logic implemented by the platforms, users are more likely to check comments once they have entered the playback page. Of the questionnaire participants, 25\% indicated that their initial impression of a song could be altered by reading a comment from another user, prompting them to leave a different comment. Furthermore, the ranking question revealed that various factors contribute to listening behavior, with basic information such as the song's title, album name, artist name, and genre having the greatest impact (94.1\% and 70.9\%, respectively). \clf{Additionally, a significant proportion of participants (78.6\%) reported that the singer's personal information also played a positive role in triggering listening behavior.} Interestingly, emotional and scene tags, as well as other people's opinions, were a priority for a quarter of the participants, inspiring new and valuable content for song suggestions. Finally, regarding the impact of user interface on listening behavior, 75\% of the participants agreed that it had an effect, provided that it was intuitive and aesthetically pleasing, and did not result in a complex and cluttered interface, as emphasized by P3.

\subsection{Design Goals}
\par Drawing upon the findings we have identified and user expectations, we have formulated the following design objectives.

\par \noindent\textbf{G.1 Use comments to help users make decisions.} In our formative study, we discovered that users take into account the comments of others when choosing songs. However, current music platforms provide inadequate support for this aspect of the user experience (\textbf{C.3}). To enhance this aspect, we propose the incorporation of a visual preview of the comments. By providing users with the ability to preview comments while browsing the song list, they can gain a general idea of the song's characteristics, which can aid them in selecting songs that align with their interests.

\par \noindent\textbf{G.2 Provide a browser-friendly way of organizing comments.} A significant number of users refrain from perusing comments on mainstream music platforms not because they are entirely indifferent to them, but due to the inadequate organization of the comments section. The current approach, which involves presenting users with a high volume of comments simultaneously, is not conducive to the exploration of comment trends (\textbf{C.2}). As such, there is an immediate need for a more effective organizational strategy that promotes efficient browsing of comments.
 
\par \noindent\textbf{G.3 Help users efficiently find comments of interest.} Users may have varying preferences for the categories of comments they wish to view, depending on their particular scenario or emotional state. However, accommodating these diverse emotional states can be challenging due to the inherently variable nature of human emotions (\textbf{C.1}). In light of this challenge, we propose the development of a system that leverages the sentiment conveyed within comments to facilitate a more seamless identification of comments that are better suited to the user's current emotional state.

\par \noindent\textbf{G.4 Intuitive and easy to use.} The questionnaire conducted among participants revealed certain limitations in the functionality of existing music streaming platforms (\textbf{C.1}). To provide an improved user experience, it is necessary to optimize our system to address these shortcomings. Additionally, it is crucial to consider both the system's usability and aesthetics since the target user base is the general public.

\section{System Design}
\subsection{Approach Overview}
\par In accordance with our design objectives, we propose a pipeline that facilitates users in effectively exploring music comments and leveraging them to make informed decisions. The system pipeline, as depicted in \autoref{fig:pipeline}, comprises a \textit{Back-end Engine} and a \textit{Front-end Visualization Interface}. The Back-end Engine preprocesses the raw music comment data to construct a corpus of comments, which is then utilized to fine-tune the pre-trained language model. Both the pre-trained and fine-tuned language models are capable of extracting sentiment, induced mechanisms, and keywords. These keywords are subsequently employed to perform topic clustering using the LDA topic model. The outputs obtained from these processes are then transmitted to the Front-end Visualization Interface. This interface features two pages that are modeled on the NetEase Cloud Music player: the first is a song list page, which presents eight keywords per song as a visual preview; and the second is a comment details page that furnishes an overview of the comments, with a chronologically segmented structure that enables users to delve deeper into the details as per their requirements.

\subsection{Back-end Engine}
\par In order to facilitate the organization and exploration of music comments, we employ language models to process the raw comment data obtained through web crawling on the \textit{NetEase Cloud Music} platform. Specifically, the language models are utilized to derive topics from the keywords extracted from each individual comment, as well as to predict the sentiment and induced mechanisms associated with the comments.

\subsubsection{Data Processing}
\par Our dataset comprises a total of $206,134,406$ music-related comments, which were collected from $3,000$ distinct songs, resulting in a cumulative data size of $280$ GB. The collected data is subjected to a three-step processing methodology. First, the relevant features of each comment message, such as the comment ID, content, and time of posting, are extracted. Second, statistical information pertaining to the comments for each song is obtained, which includes data on the number of comments received during different time periods, listener demographics, and geographic locations. Third, we manually labeled $8000$ comments for the purpose of fine-tuning the language model in the prediction task. A summary of the key statistics of our dataset is presented in \autoref{tab:statistics}.

\begin{wraptable}{r}{5cm}
    \small
    \centering
    \caption{Statistical information of the corpus.}
    \label{tab:statistics}
    \begin{tabular}{|c|c|}
		\hline
		\textbf{Max Length} & 280  \\
            \hline
		\textbf{Avg Length} & 21.41  \\
            \hline
		\textbf{Vocabulary Size} & 419,201     \\
            \hline
		\textbf{ Token \#} & 5,819,300,217   \\
            \hline
		\textbf{Sentence \#} & 29,367,459    \\
            \hline
		\textbf{Sentence with emoticon \# } & 19,382,522\\
            \hline
    \end{tabular}
\end{wraptable}


\subsubsection{Contextual Keywords and Topic Detection}
\par In this study, we utilize a self-supervised method~\cite{grootendorst2020keybert} for the purpose of labeling keywords based on their contextual relevance to the sentence. This is achieved by leveraging the pre-trained Language model and the feature vector of the sentence~\cite{vaswani2017attention}, which can be used to learn the context of the sentence. The embedding representations of the keywords are obtained, and the cosine similarity between the keyword and the sentence is measured. If the similarity value exceeds a certain threshold, the word is marked as a keyword. In our approach, we select the top $5$ words with the highest similarity as the keywords of the sentences. To further enhance the analysis of the collected music comment dataset, we apply the eLDA (Ensemble Latent Dirichlet Allocation) topic model~\cite{brigl2018extracting}. We train a collection of $8$ topic models, each of which extracts a cluster of $20$ topics. The topic clusters are subsequently grouped by DBSCAN~\cite{Ester:1996:DBSCAN}.


\subsubsection{Sentiment and Induced Mechanism Prediction}

\begin{wraptable}{r}{6cm}
    \small
    \caption{Experimental results in the sentiment prediction task.}
    \label{tab:result}
    \begin{tabular}{ccccc}
		\hline
		\multirow{2}{*}{\textbf{Models}} & \multicolumn{2}{c}{\textbf{Train}} & \multicolumn{2}{c}{\textbf{Test}} \\
		& \textbf{EM}   & \textbf{F1}   & \textbf{EM}   & \textbf{F1}    \\ \hline
		SVM              & 59.8 & 61.4 & 59.2 & 60.3  \\
		BiLSTM           & 65.2 & 73.1 & 69.8 & 72.4  \\
		Chinese-BERT-wwm & 66.3 & 85.6 & 70.5 & 87.7  \\
		RoBERTa-wwm-ext  & 67.4 & 87.2 & 72.6 & 89.4  \\
		\textbf{Ours}     & \textbf{68.5} & \textbf{88.4} & \textbf{74.2} & \textbf{90.6}  \\ \hline
    \end{tabular}
\end{wraptable}

\par In this study, we have chosen the pre-trained language model \textit{RoBERTa-wwm-ext-large, Chinese}~\cite{cui-etal-2020-revisiting} as it demonstrates favorable suitability for Chinese language semantics and exhibits commendable performance across various NLP tasks. Our research encompasses two classification tasks, specifically geared towards predicting the sentiment and induced mechanism of music comments. To accomplish these tasks, we develop two separate language models that undergo fine-tuning. Regrettably, no substantial publicly available dataset of labeled Chinese musical comments, pertaining to sentiment or induced mechanism, exists. Consequently, we manually label the collected data, consisting of approximately $10,000$ music comments, for the sentiment prediction task. \clf{Following this, the labeled data is divided into the training set, validation set, and test set in a $6:2:2$ ratio. These subsets are utilized for fine-tuning the pre-trained model. During the fine-tuning process, we load the initial weights of \textit{RoBERTa-wwm-ext-large, Chinese} using the \textit{HuggingFace Transformers} framework~\cite{wolf2020huggingfaces}. We adopt an Adam optimizer with a learning rate of $5$e-$5$ and employ sparse categorical entropy as the loss function. The fine-tuning is concluded at the epoch where the model achieves optimal performance on the validation set. We also apply similar fine-tuning and evaluation procedures to other commonly used models, comparing their performance against our approach in terms of EM~\footnote{EM (Exact Match) calculates the percentage of ground truth answers that match the predicted outcomes.} and F1 metrics. A comprehensive presentation of the obtained results is provided in \autoref{tab:result}.}


\par In the induced mechanism prediction task, we meticulously label a dataset comprising $8,000$ music comments. Notably, the distribution of labels across the four induced mechanism categories, namely \textit{Music Evaluation}, \textit{Personal Memory and Experience}, \textit{Background Information}, and \textit{Others}, is approximately equal, with a ratio of $1:1:1:1$. \clf{These categories have been derived and summarized based on the framework proposed by Juslin~\cite{JUSLIN:2013:everyday}, and their respective meanings are as follows: 1) \textit{Music Evaluation}: Comments falling into this category pertain to the lyrics, melody, rhythm, performance style, recording mode, and production team (excluding the singer). They reflect the listeners' immediate attitudes, evoke visual imagery, and encompass aesthetic judgments regarding the quality of the music clip. 2) \textit{Personal Memories and Experiences}: Within this category, comments are triggered by a song and evoke personal memories or experiences of the listeners associated with certain events in their lives. 3) \textit{Contextual Information}: Comments in this category extend beyond the song itself and encompass broader social events or human knowledge that may be relevant to a wider audience. They offer contextual information alongside the listener's perspective on the song. 4) \textit{Others}: This category encompasses music comments that are either incomprehensible and philosophical or involve listeners simply copying and pasting lyrics they enjoy. Such comments cannot be effectively classified into any of the aforementioned three categories. Through this labeling process, we establish a comprehensive dataset that accounts for the diverse induced mechanisms present in music comments.}

\subsection{Front-end Visualization}
\par The primary principle of our design is to leverage and enhance familiar visual metaphors. We adhere to the guiding principle of ``overview first, zoom and filter, then details-on-demand''~\cite{shneiderman1996eyes} for users to actively explore music comments, thereby enhancing their music listening experience. In this section, we will present the front-end interface of our system, comprising two distinct pages. Subsequently, we will introduce the visual encodings and the successive stages involved in the reconstruction of comments using a map metaphor-based approach.

\subsubsection{Song List Page}

\begin{figure}[h]
 \centering 
 \includegraphics[width=\linewidth]{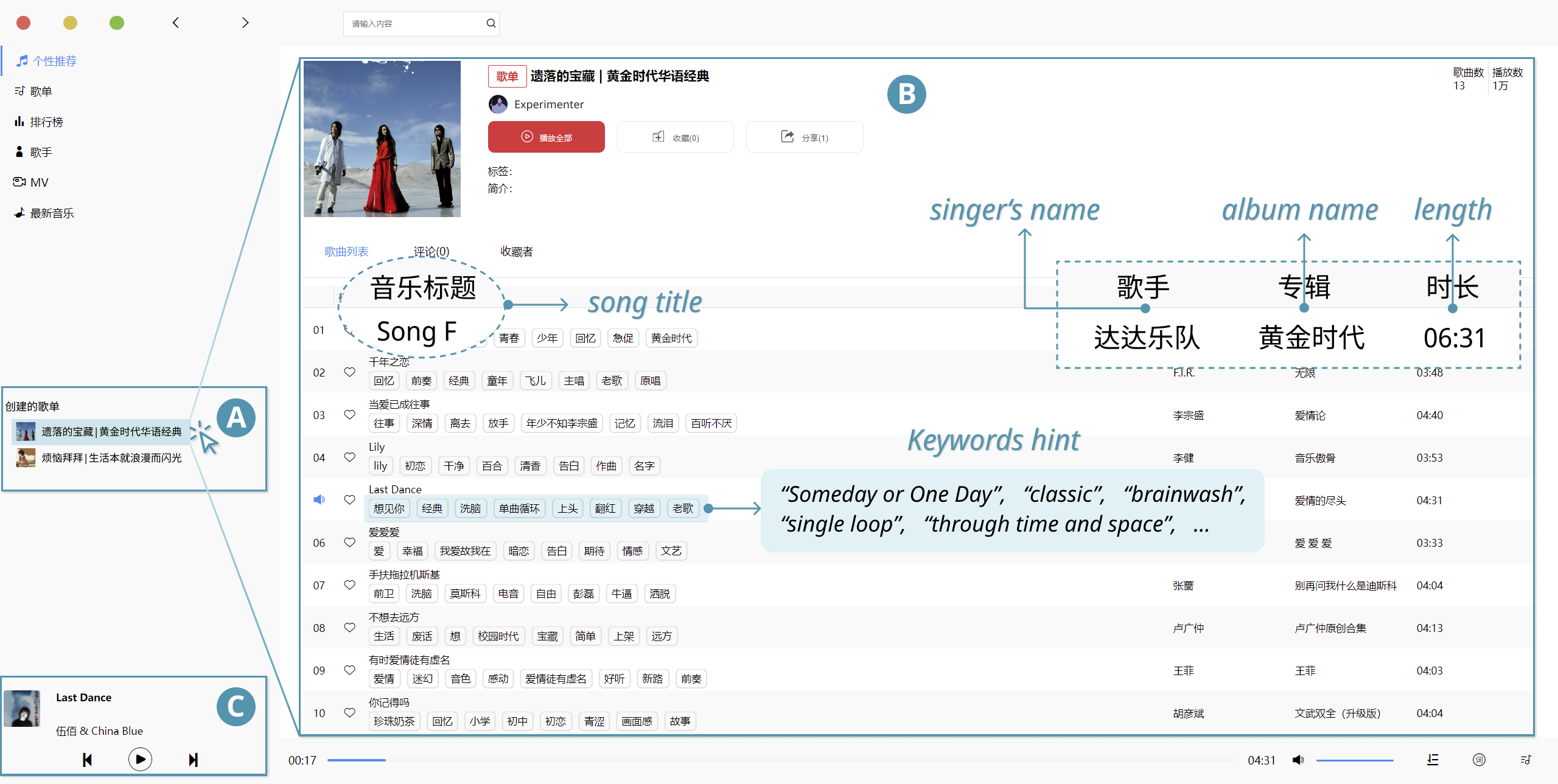}
 \caption{Song List Page. (A) Song List Library View, (B) Selected Song List View, and (C) Music Player View. Each song in the Song List view has eight keywords hints extracted from music comments.}
 \label{fig:songList}
\end{figure}

\par The Song List Page comprises three primary views, which are illustrated in \autoref{fig:songList}. First, the Song List Library View presents multiple alternative song lists for users to browse. Second, the Selected Song List View displays comprehensive details about the songs in the selected song list, such as the song title, singer's name, album name, and eight high-frequency keywords extracted from music comments, presented as visual preview tags. Third, the Music Player View enables users to pause, resume, switch the currently playing song, and click the album cover image to navigate to the Comment Details Page. This study aims to investigate the impact of incorporating visual preview tags on users' song selection behavior, and consequently, the entire layout of the interface closely adheres to \textit{NetEase Cloud Music}.

\subsubsection{Comment Details Page}

\begin{figure}[h]
 \centering 
 \includegraphics[width=\linewidth]{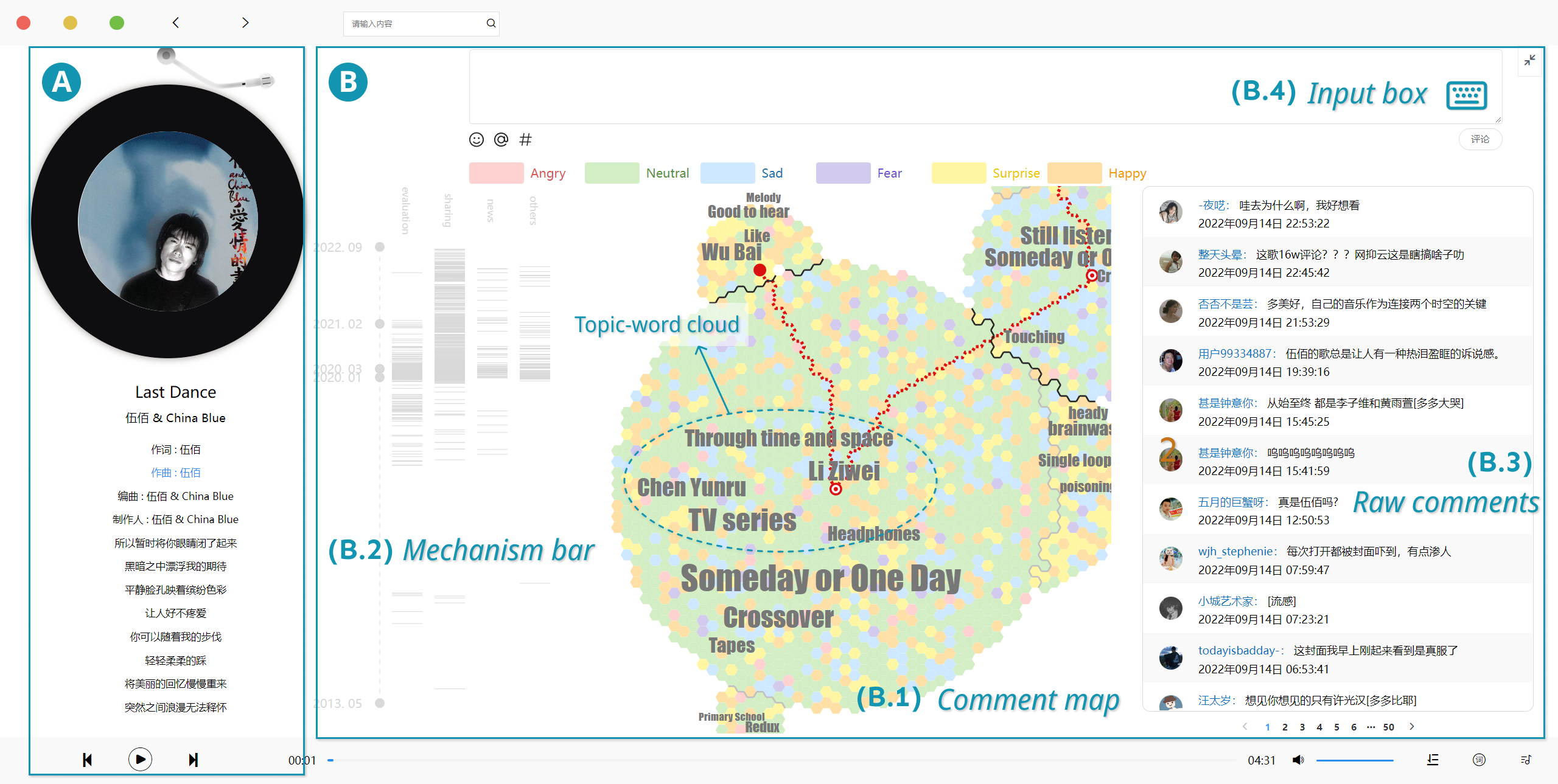}
 \caption{Comment Details Page. (A) Lyric View, and (B) Comment View. Inside the Comment View, there is (B.1) a comment map in the middle, the topic-word cloud above the comment map summarizes all the topics, (B.2) four induced mechanism bars with a timeline on the left, (B.3) raw comment content on the right, and (B.4) an input box on the top.}
 \label{fig:CommentDetails}
\end{figure}

\begin{figure}[h]
 \centering 
 \includegraphics[width=\linewidth]{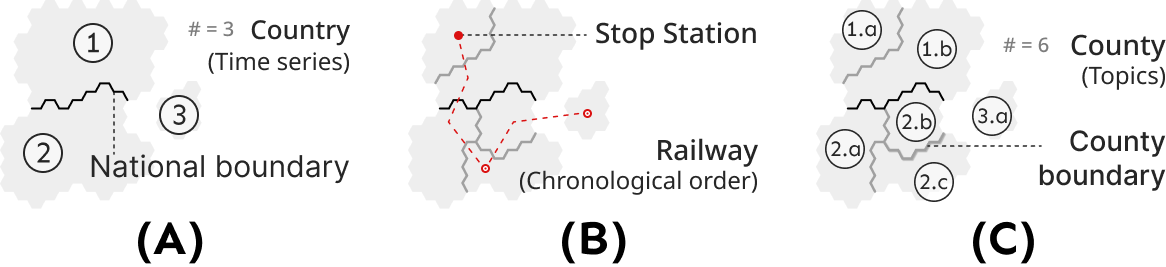}
 \caption{A legend explaining our map metaphors. (A) Country corresponds to the group of comments in the same period. (B) The center of each country has a stop station marked by a red circle. And the railway crosses all the countries' stop stations. (C) Each topic within a specific time period is associated with a county.}
 \label{fig:legend}
\end{figure}

\begin{figure*}[h]
 \centering 
 \includegraphics[width=\linewidth]{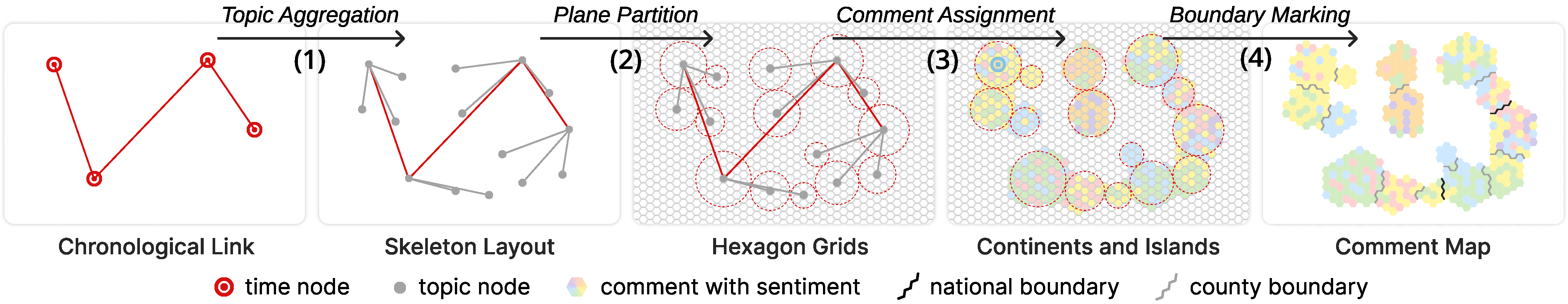}
 \caption{Comment map construction. (1) Extract topics in different time periods to generate a skeleton layout. (2) Split the skeleton layout into hexagonal grids using Voronoi diagrams. (3) Comments are assigned plate cells to generate a map layout, and (4) complementary boundaries are added to separate different time periods and topic clusters.}
 \label{fig:maplayout}
\end{figure*}

\par As depicted in \autoref{fig:CommentDetails}, the Comment Details Page comprises two primary views. First, the Lyric View (\autoref{fig:CommentDetails} A), situated on the left, serves as a music player that displays lyrics. Second, the Comment View, located on the right, integrates the comment browsing (\autoref{fig:CommentDetails} B.1 B.2 B.3) and comment posting modules (\autoref{fig:CommentDetails} B.4), enabling the user to selectively view the text of the original comment in both abstract and sequential formats. Specifically, we propose a map-metaphor-based comment overview design (\autoref{fig:CommentDetails} B.1) to graphically organize comment texts as countries, counties, and railways (\autoref{fig:legend}). The features of music comments are mapped onto various visual elements, which we present in a sequence from the entire to the partial view.

\par \noindent\textbf{Time Period.} Each comment published is accompanied by a timestamp, and considering the social nature of music in contemporary society, various events such as movie releases, TV shows, live performances, or news about music artists can significantly influence people's perceptions of songs, leading to an upsurge in published comments. To effectively capture this cyclic clustering phenomenon, we employ a time-segmentation algorithm that determines the optimal cut point based on the growth of comments, thereby dividing the time series into multiple distinct periods. \clf{Specifically, our approach adopts a top-down methodology inspired by Keogh and Pazzani~\cite{keogh2004segmenting}, recursively segmenting the time series until a predefined stopping condition is met. The algorithm evaluates all possible divisions of the time series, identifying the most suitable position for splitting. The resulting subsequences are then tested to verify if their approximation error falls below a specified threshold. If the error exceeds the threshold, the algorithm proceeds to recursively segment the subsequences until all segments achieve approximation errors below the threshold.} As depicted in \autoref{fig:legend}(A), we define a ``country'' as a group of comments within the same time period. Furthermore, in \autoref{fig:legend}(B), we illustrate the center of each ``country'' denoted by a stop station represented by a double concentric circle, except for the station corresponding to the earliest time period, which is marked with a solid circle. To convey the temporal order of these time periods associated with the countries, we employ a ``railway'' metaphor represented by a red stroke traversing the centers of all countries. Additionally, if there are bordering relationships between countries, we emphasize the boundaries between them using black strokes.

\par \noindent\textbf{Topic.} Through unsupervised methods, several topics can be derived from comments within a specific time period. In our visualization, each topic is associated with a ``county'' (\autoref{fig:legend}(C)), which serves as a subunit of the country. The layout algorithm determines whether counties should be adjacent or not. If a county shares borders with other counties, we employ dark gray strokes to demarcate their boundaries, allowing users to distinguish between topics. Moreover, word clouds are utilized to represent topics through keywords, aiding users in gaining an intuitive understanding of their content. The size of the keywords reflects their prevalence in the topic; the larger the size, the more frequently it occurs in the comments under that topic.

\par \noindent\textbf{Sentiment.} Each comment is represented as a puzzle piece, and these puzzle cells combine to form the territory of the country. The corresponding puzzle pieces are assigned one of six colors based on the sentiment class labels predicted by our back-end model, following the color-sentiment mapping relationship presented in~\cite{hanada:2018:correspondence}. Specifically, the mapping is as follows: \textit{{Happy -- Orange}}, \textit{{Angry -- Red}}, \textit{{Sad -- Blue}}, \textit{{Fear -- Violet}}, \textit{{Surprise -- Yellow}}, and \textit{{Neutral -- Green}}. Additionally, users can obtain the raw contextual information related to a comment by clicking on its cell, enabling them to better understand the sentiment conveyed.

\par \noindent\textbf{Induced Mechanism.} The underlying mechanism that motivates users to write music comments is largely influenced by the semantic content of the comments. Therefore, we leverage this aspect to provide users with insight into a particular topic. The four mechanisms are visually represented as light gray stripes (\autoref{fig:CommentDetails} B.2) that run parallel to the vertical timeline on the left. If a user selects a particular county associated with a topic, the corresponding mechanism is highlighted with dark gray, providing a cue to the user.

\subsubsection{Map Construction}

\par The generation of a comment map entails the initial identification of a chronological link comprising a sequence of time nodes. These time nodes are defined as the cut points derived from the application of a time segmentation algorithm to the comments. Once the chronological link has been established, a subsequent four-step process, as depicted in \autoref{fig:maplayout}, is required to produce the final comment map.

\par \noindent\textbf{Topic Aggregation.} An unsupervised learning approach is employed to extract topics from the comments associated with each time node. Subsequently, each extracted topic is added to the corresponding time node as a child node, referred to as the ``topic node'', whose radius is determined by the square root of the number of comments it contains. Once all topic nodes have been added, we replace each time node with its child node possessing the largest radius. This yields a topic tree, upon which the bubble tree drawing algorithm~\cite{grivet2006bubble} is applied to produce the skeleton layout.

\par \noindent\textbf{Plane Partition.} After the generation of the skeleton layout, a set of hexagons is overlaid within its bounding box. The Voronoi Diagram is used to divide these hexagons. The seed points of the Voronoi Diagram are projected as the center points of the hexagons into the bounding box, where the number of seed points, denoted as $N_s$, is determined as $N_s := \frac{W \times H}{\pi}$, where $W$ and $H$ are the width and height of the bounding box, respectively. This design is intended to make the number of hexagons per unit area approximately equal to the number of comments for the following operations. Alternatively, other types of shapes embedded in the plane, such as squares and triangles, have been considered. However, using squares may cause a distinct jaggedness that leads to unaesthetic effects, while using triangles requires dividing them into upper and lower triangles, and encoding the same information using two basic shapes may lead to confusion in understanding. In contrast, the hexagonal layout is a better choice.

\par \noindent\textbf{Comment Assignment.} In this stage, we proceed to assign a distinct hexagonal grid to each comment. We adopt a sequential approach, starting from the center point of each topic node and assigning comments in chronological order. A comment will seek out an unassigned hexagonal grid closest to the center point. Once all comments have been assigned, the skeleton layout transforms into a map layout.

\par \noindent\textbf{Boundary Marking.} Upon generating the map layout, it is necessary to establish boundaries to delimit the visual metaphors depicted in the map. To this end, we employ black strokes, referred to as ``national boundaries'', to separate topic plates in adjacent time series. Additionally, gray strokes, or ``county boundaries'', are employed to separate nearby topic plates in the same time series. The national boundaries and county lines serve to distinguish the different plates visually. In instances where closed plates in different time series or different topics are not adjacent to one another, the national boundaries and county lines are not distinctly marked. The resulting gaps between these boundaries can aid users in distinguishing between them.

\section{Usage Scenario}
\par We demonstrate the effectiveness of our approach in exploring song comments by presenting a usage scenario based on actual music comments extracted from \textit{NetEase CloudMusic}.

\par
\begin{figure}[h]
\centering 
    \includegraphics[width=\linewidth]{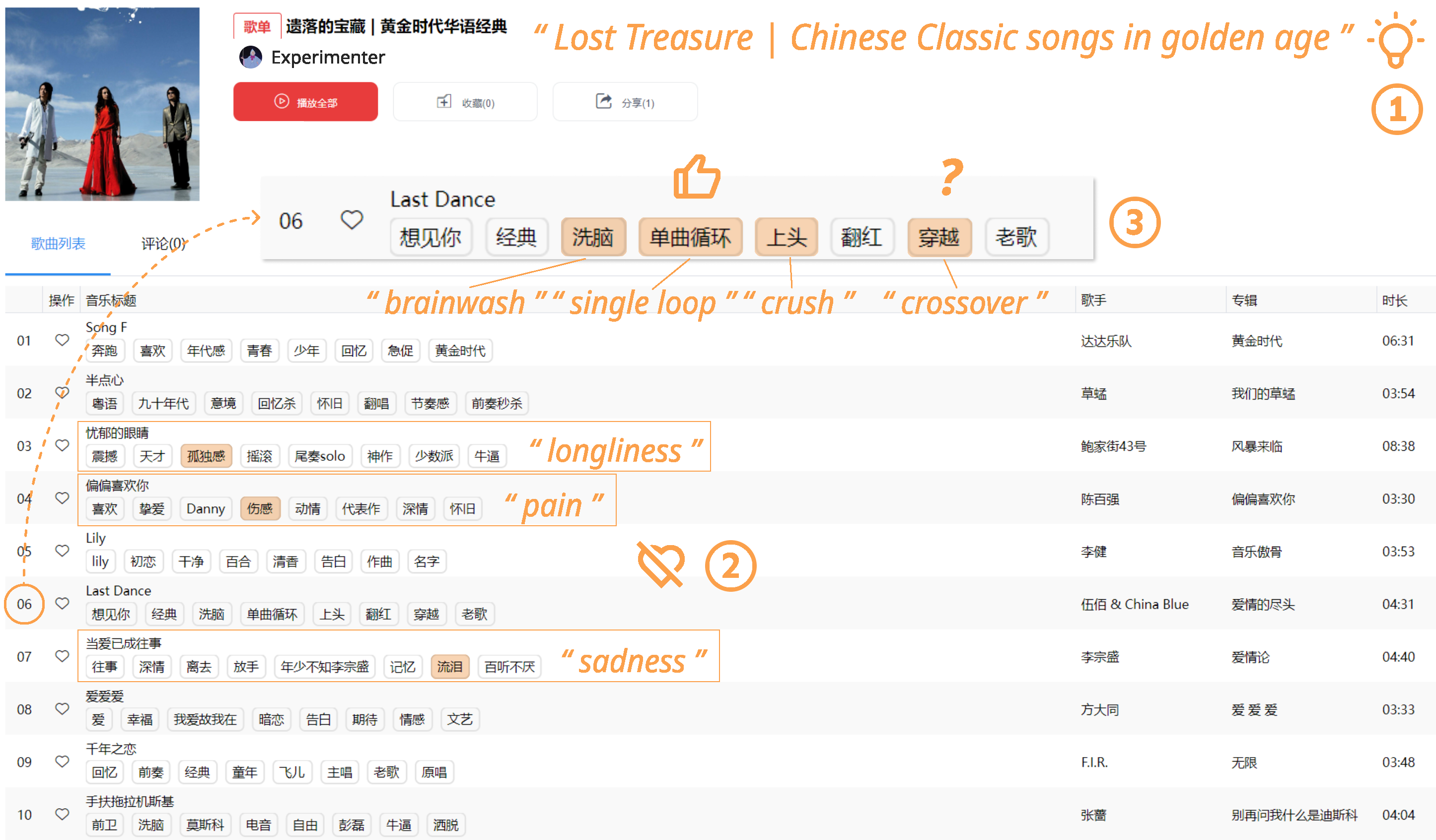} 
    \caption{Usage scenario: Music preview tags bring surprises. (1) The user enters a song list consisting of Chinese classic songs, (2) negative emotional tags make the user not want to listen to certain songs, and (3) some interesting tags attract the user.} 
    \label{fig:case1} 
\end{figure}

 Alice, an oldies fan, selected a song list of 13 Chinese classical songs. As shown in \autoref{fig:case1}, after observing the tags of several songs on the song list page, she chose ``Last Dance'' because it had tags such as ``brainwash'', ``crush'', and ``single loop''. Alice found these tags to be indicative of a song that people would want to listen to repeatedly. She also had questions about some tags, such as ``crossover'' and ``someday or one day''. She double-clicked on the song and navigated to the comment details page. Alice examined the timeline in the comment view, which is divided into four stages, and then moved to the comment map to get an overview of the comments corresponding to these stages. She quickly identified the earliest comment area and discovered that early comments were about songwriters and melodies, supported by the mechanical bars on the left that are mainly concentrated in ``evaluation''. By examining the color of the cells in this area, Alice determined that the early comments were overwhelmingly positive.

\begin{figure}[h] \centering 
    \includegraphics[width=\linewidth]{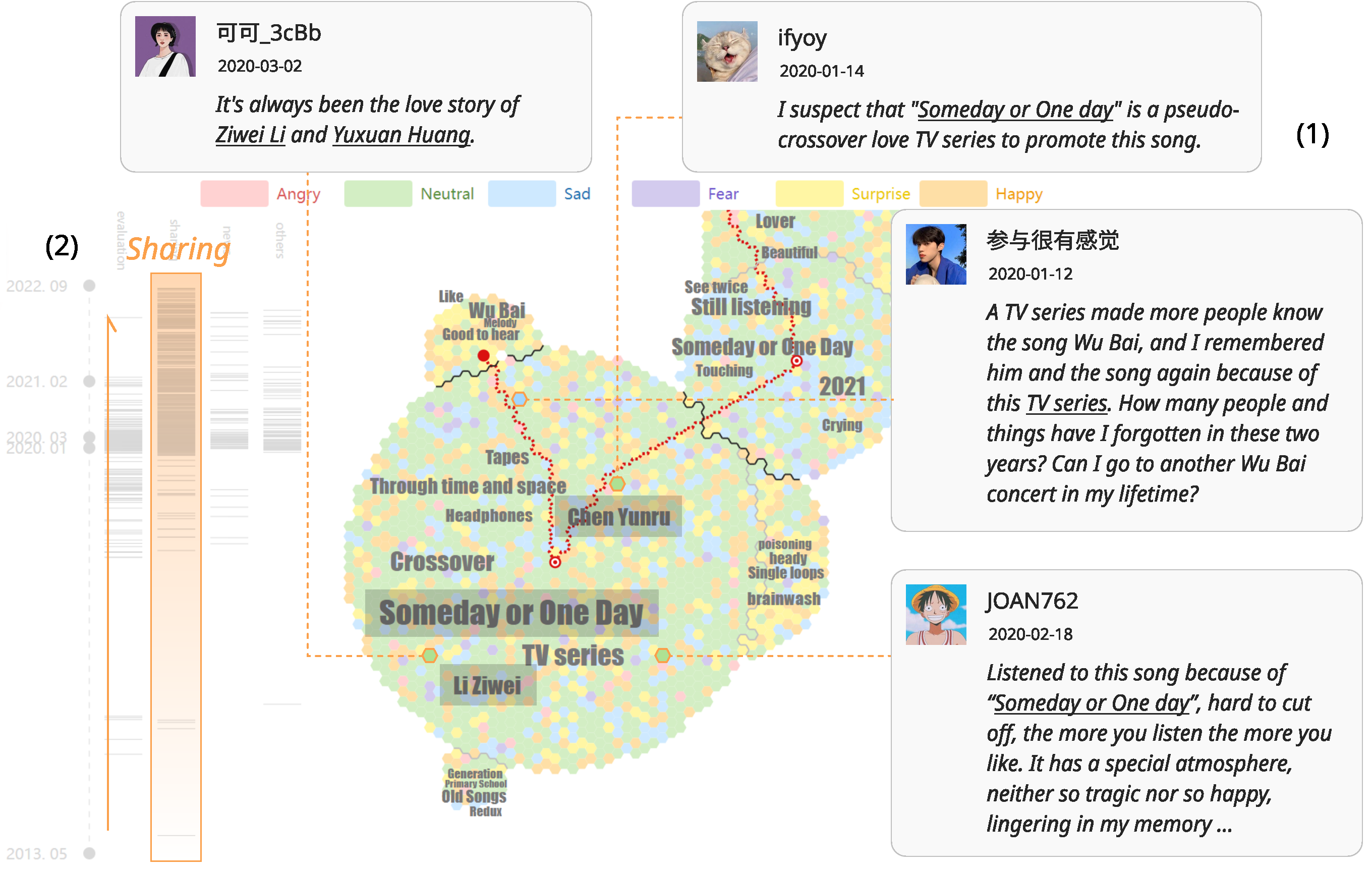} 
    \caption{Usage Scenario: Impact of external events on song comment topics. (1) Comments help the user understand the keywords in word clouds representing topics, and (2) the comments induced by \textit{Sharing} mechanism become more and more common over time.} 
    \label{fig:case2} 
\end{figure}

\par Subsequently, Alice's attention was captured by the second area with the red ``railway'', which had the highest number of comments during this period, following her exploration of the first area. As the second time period on the left timeline spanned from January to March 2020, she was surprised to find that a considerable number of comments were generated in only three months, far more than in the previous seven years. This timeframe contained three topics, and she decided to view the largest one first. However, she noticed the presence of incomprehensible keywords (e.g., ``Someday or One Day'', ``Li Ziwei'', ``Chen Yunru'') and clicked on some comments to see the original text. As depicted in \autoref{fig:case2} (1), some comments helped her grasp the meaning of these words: ``Someday or One Day'' is a TV series about time travel and love, while ``Li Ziwei'' and ``Chen Yunru'' are the characters in the TV series, which features the song. This resolved her previous confusion about the reasons for the surge in comments - a popular TV series drew attention to the song. Alice also observed that the cells during this period exhibited a considerable amount of green, with some blue added, indicating that the sentiment behind the comments was neutral to sad. Upon clicking on these comments, she realized that they were mostly discussing the TV show rather than the song itself. Moreover, during the same period, comments expressing \textit{surprise} and \textit{happiness} were primarily related to the songs rather than external events. Alice then investigated the third and fourth time periods and found that the dominant sentiment remained neutral. Additionally, as depicted in \autoref{fig:case2} (2), comments triggered by the \textit{Sharing} feature became increasingly prevalent over time. After concluding her analysis, she remarked that ``\textit{the song itself can elicit positive sentiments, but external events can influence the sentiment trend, causing the discussion to shift away from the song and towards neutral or sad sentiments.}'' Furthermore, she added that ``\textit{it is challenging to capture this change in the original comment section of existing music streaming platforms.}''

\begin{figure}[h] \centering 
    \includegraphics[width=\linewidth]{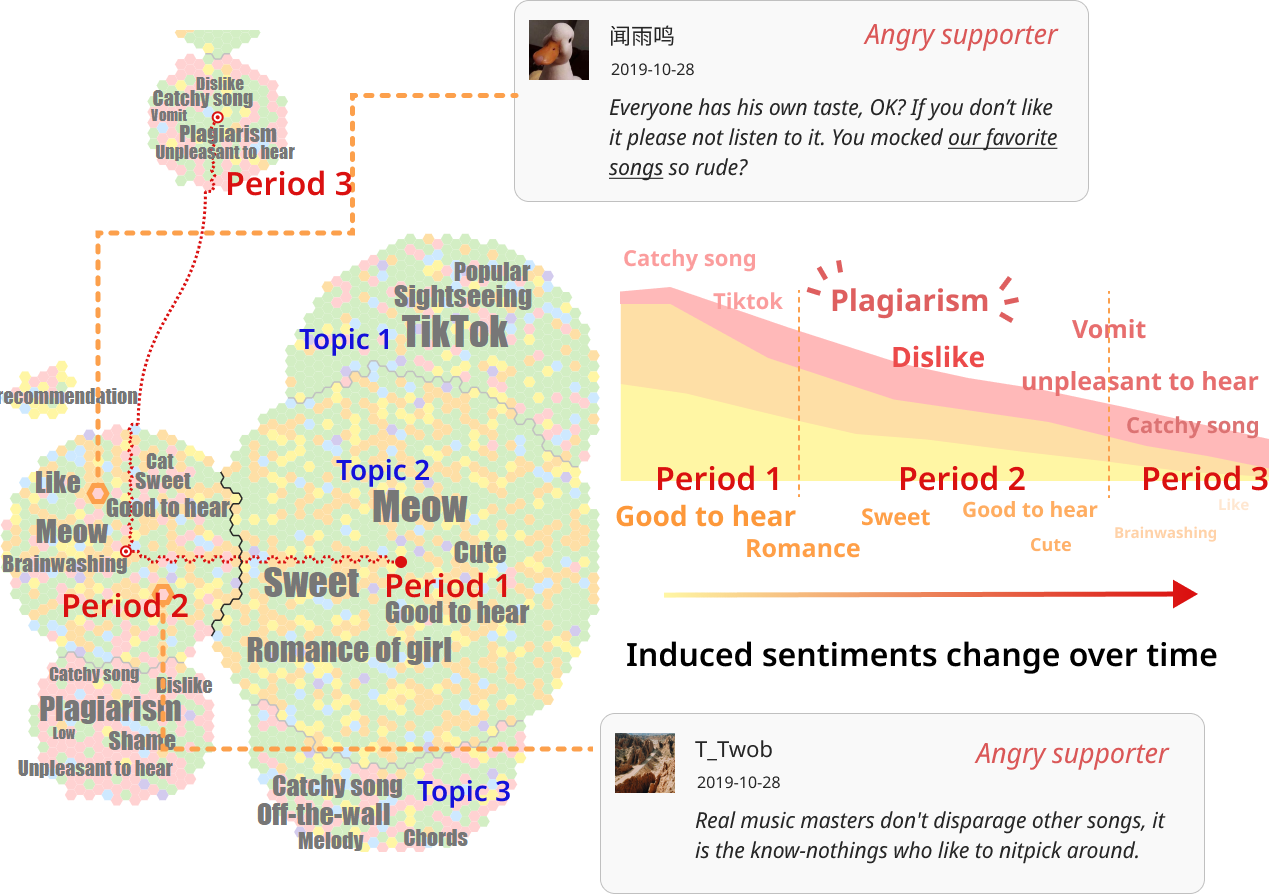} 
    \caption{Usage scenario: Changes in the induced sentiment direction of song comments over time.} 
    \label{fig: case3} 
\end{figure}

\par Alice's curiosity piqued as she ventured to explore another song named ``Say Meow Meow'', a controversial and popular tune on the internet in recent years, with divergent opinions among its listeners. She intended to employ our system to investigate how the comments on this song have evolved over time. As depicted in \autoref{fig: case3}, in the first period, the listeners of the first topic generally expressed favorable opinions towards the song, describing it as nice, sweet, and romantic, evoking emotions of \textit{surprise} and \textit{happy}. The second topic pertained to the short video platform \textit{TikTok}, where several listeners stated hearing the song on this platform before coming to the music streaming platform, with the sentiment being \textit{neutral}. However, some \textit{angry} comments expressed a dislike for \textit{TikTok}. The third topic centered on \textit{angry} sentiments, and the keywords of the topic suggested that these comments criticized the song as an off-the-wall catchy tune. Alice observed that the topic parallels for different time periods were the same, signifying the two groups of listeners who liked and disliked the song. The color of the cells in this area indicated that even among those who liked the song, there were also \textit{anger}, which were mainly a response to negative comments. Conversely, the group that disliked the song deemed it a catchy tune but used plagiarism as a potent weapon to criticize it. Most of the negative comments concerning the song were about its alleged similarity to another tune. In the third period, the sentiment trend was overwhelmingly negative where red and green cells dominated, with few positive sentiments, and most negative comments were still related to plagiarism. Alice deduced that the comments on a popular song's sentiment changed over time. After its release, two major groups, namely supporters and opponents, emerged rapidly. Initially, the supporters outnumbered the opponents, with the induced sentiment showing a positive trend. However, over time, the opponents accused the song of being plagiarized, which gradually reversed the listeners' perception of the song, and the induced emotions turned negative.

\section{User Study}
\par In order to evaluate our proposed augmented music application, we formulated two research questions as follows: \textbf{\textit{RQ1: To what extent is the augmented music application usable and effective in an actual listening process?}} and \textbf{\textit{RQ2: How do users interact with and respond to this application during real-world listening sessions?}} To address these questions, we conducted a user study under controlled laboratory conditions where participants utilized our enhanced music app and the baseline music application, i.e., \textit{NetEase Cloud Music}.

\subsection{Participants}
\par A total of 20 participants (designated as P1 -- P20) were enlisted for our study. Of these participants, 14 were males and 6 were females, with a mean age of 22.6 years (SD = 1.47). Participants were assigned to one of four groups, with five participants randomly assigned to each group, following the formative study.


\subsection{Procedure}
\par We evaluated our improved music application against the baseline by conducting a comparison. To reduce the impact of ordering effects, participants were asked to explore two song lists in a counterbalanced order using two music apps. The song lists were designed to resemble existing song lists on music streaming platforms, with songs in the same list sharing a common theme. For instance, one song list contained classic songs, while the other featured songs that evoke happiness. We ensured that all songs had a minimum of hundreds of comments, but none were excessively popular, to avoid potential biases. Prior to the tasks, we provided participants with instructions on how to use both music apps and allowed them five minutes to familiarize themselves with each app. Participants completed two tasks for each app. The first task was to \textbf{identify three songs in a specific list that matched three detailed descriptions}, while the second task was to \textbf{identify the key features of comments of the designated songs}, and participants could listen to any music in the list while completing the task. After each exploration round, participants were required to complete a 5-point Likert scale in-task questionnaire to gather feedback on the \textit{usability}, \textit{effectiveness}, and \textit{user experience} of the music app. The questionnaire included options ranging from strongly disagree (1) to strongly agree (5). Finally, we conducted semi-structured interviews with participants, consisting of several open-ended questions.

\subsection{Results}
\par Our study employs both quantitative and qualitative methods to analyze the data. The quantitative analysis involves the use of the Friedman test with post hoc Wilcoxon signed rank test, and descriptive statistics are employed to summarize the responses to the in-task questionnaire.


\subsubsection{RQ1: How usable and effective is the augmented music application in a real listening process?}
\par We investigated the usability of our proposed music app experienced by participants compared to the baseline app. The Friedman tests showed that there are significant differences in terms of \textit{Helpfulness in music exploration} ($\chi^2(2)=18.968, p<0.0001$), \textit{Helpfulness in comment browsing} ($\chi^2(2)=33.641, p<0.0001$), \textit{Satisfaction with the visual representation} ($\chi^2(2)=30.349, p<0.0001$), \textit{Likelihood for future use} ($\chi^2(2)=21.076, p<0.0001$). Apparently, our music app had a better visual representation of comments and helped users enjoy more comments of interest ($AVG=4.500, SD=0.761; AVG=4.400, SD=0.821$) than the baseline ($AVG=2.200, SD=1.005; AVG=1.800, SD=0.616$). We also investigated how our music apps can enhance the user experience in different aspects. Both apps demonstrate significant differences in terms of \textit{Improving pre-impression} ($\chi^2(2)=24.479, p<0.0001$), \textit{Making it easier to have an overview of comments} ($\chi^2(2)=30.857, p<0.0001$), \textit{Establishing connections with other listeners} ($\chi^2(2)=14.862, p=0.0007<0.001$), \textit{Making sentiments of comments easier to understand} ($\chi^2(2)=36.308, p<0.0001$), \textit{Clarifying the chronological order of the comments} ($\chi^2(2)=26.078, p<0.0001$), and \textit{Enhancing engagement of comments} ($\chi^2(2)=15.741, p<0.0001$). Our app was more effective in obtaining an overview of comments, understanding sentiment, and clarifying the temporal order of comments ($AVG=4.600, SD=0.754; AVG=4.400, SD=0.940; AVG=4.500, SD=0.761$) than the baseline ($AVG=1.600, SD=0.940; AVG=1.500, SD=0.688; AVG=1.750, SD=1.118$). These results are consistent with qualitative feedback from participants in the interviews. Our app was considered more usable and helpful than the baseline because our app ``\textit{reduces the time wasted on aimless browsing}'' (P2), ``\textit{increases the exposure of comments of multiple contents}'' (P5), ``\textit{filters uninteresting information}'' (P9), ``\textit{provides an attractive visual design for boring text messages}'' (P17). The proposed visual metaphors were ``\textit{easy to understand with a beginner's guide}'' (P3).



\subsubsection{RQ2: How will users interact with and be influenced by this application during real listening sessions?}
\par \noindent\textbf{Users performed fewer ``click to listen'' operations and complete searching tasks more quickly.} Music search can be time-consuming, especially when users are presented with numerous similar songs. Tags can streamline the search process and help users find relevant songs more quickly by allowing them to narrow down their search based on their interests and preferences. This can significantly reduce the number of songs that need to be considered. Our app's tags were reported to have allowed a user (M6) to exclude songs belonging to genres that did not align with their tastes.

\par \noindent\textbf{Users got more information from the comment map, even though they actually viewed fewer comments.} In Task 2, participants had to identify and summarize two main topics from the comments section using the baseline app, but had difficulty due to the large number of comments and had to spend more time manually searching for relevant information. Our app improved efficiency by providing the entire comment timeline and semantically similar groups of comments, resulting in improved accuracy and ease of summarization.

\par \noindent\textbf{Users felt more positive about their listening and browsing experience with our app.} The two music applications were found to significantly differ in terms of users' emotional experiences, including happiness ($\chi^2(2)=21.283, p=0.000<0.01$), exhaustion ($\chi^2(2)=13.406, p=0.009<0.01$), hopefulness ($\chi^2(2)=22.085, p=0.000<0.01$), relaxation ($\chi^2(2)=18.269, p=0.001<0.01$), and satisfaction ($\chi^2(2)=23.677, p=0.000<0.01$). Our application elicited higher levels of happiness ($t=-5.586, p=0.000<0.01$), hopefulness ($t=-6.749, p=0.000<0.01$), satisfaction ($t=-6.631, p=0.000<0.01$), and lower levels of exhaustion ($t=4.196, p=0.000<0.01$). Furthermore, users could easily empathize with other listeners and were more attracted to the comments ($t=-4.700, p=0.000<0.01; t=-4.342, p=0.000<0.01$). However, no significant differences were found in terms of feeling dysphoric or overwhelmed.

\par The results of the Friedman tests indicated significant differences between the two music apps with regards to users' workload on selecting music ($\chi^2(2)=26.619, p=0.000<0.01$), workload on understanding comments ($\chi^2(2)=26.286, p=0.000<0.01$), and attention load ($\chi^2(2)=9.740, p=0.045<0.05$). Specifically, our music app demonstrated a noteworthy reduction in workload in both selecting music and understanding comments ($AVG=2.050, SD=0.826; AVG=2.100, SD=0.968$) when compared to the baseline app ($AVG=4.050, SD=0.826; AVG=4.300, SD=0.923$). However, no significant differences in cognitive load were observed.

\par \noindent\textbf{Categorizing comments logically can reduce the workload of users.} Traditional comment organization poses difficulties in identifying and organizing relevant comments on specific topics as commenters may participate in multiple threads, leading to a shortage of comments related to particular topics. Some participants in our study found it challenging to comprehend the significance of comments related to Ziwei Li due to this approach. However, users of our comment map did not experience such confusion.

\par  \noindent\textbf{The attentional and cognitive load of users is negligible.} Most participants (18 out of 20) found the cognitive load associated with the keyword cues to be low since users only need to attend to eight words per song, which is faster than listening to song clips. The new visualization also reduces cognitive load. A few participants (P2, P4, P14, and P20) usually browse only the most recent comments, but they showed interest in exploring more due to the appealing nature of the comment map, according to P14.

\par \noindent\textbf{The additional information forced users to think about what they really wanted, which introduced a new workload.} Participants showed interest in the tags and expressed concern for their preferences during exploration. P6 found that tags helped them identify the music they want to listen to, while P8 commented on the effort the tags required. P12 mentioned the workload was heavy but worth it to discover more songs they liked.

\section{Discussion and Limitation}

\par \noindent\textbf{Comment Specificity in Streaming Music Platforms.} Comments on streaming music platforms like \textit{YouTube Music} and \textit{NetEase Cloud Music} focus more on music rather than overall user experience. The informality of comments and lack of text makes data mining difficult~\cite{song2014short}. We used self-tagged data to adapt a pre-trained language model to achieve some accuracy, but accurate capture of sentiment and induced mechanisms in comment texts still require training on a larger corpus. Music comments on social media or online forums may be more centered on user experience. This study aims to improve browsing efficiency on streaming music platforms using a map metaphor.


\par \noindent\textbf{Visual Design and Learning Curve.} The design of our system followed a user-centered process. We selected a popular streaming music platform, \textit{NetEase Cloud Music}, as the baseline for our study. To identify the limitations of the platform and the challenges users faced when exploring song comments, we conducted a formative study with 104 participants. During the design process, we prioritized ease of use and aesthetics, employing intuitive visual metaphors such as countries, counties and railroads to represent comments. Since our system was built on top of an existing music platform, we aimed to maintain interface consistency with the baseline platform as much as possible. Once introduced and explained, users were able to quickly adapt to our visualizations and explore the system with ease.

\par \noindent\textbf{Generalizability and Scalability.} Regarding generalizability, our visual design has potential applications in various scenarios involving the viewing of comment text, such as video comments, social media comments, and product comments. The back-end model is also capable of processing different types of text inputs; however, it requires tuning to better suit the characteristics of the new corpus for improved performance. In terms of scalability, the front-end visualization is not capable of handling vast amounts of comment data. The back-end engine used for detecting sentiments, induced mechanisms, and keywords utilizes a typical machine learning process that can be adapted to other data types with only retraining. However, as our approach uses supervised machine learning, it may face challenges in obtaining sufficient labeled training data when targeting other specific tasks.

\par \noindent\textbf{Limitations.} First, while our approach facilitates a seamless transition for users from a comprehensive overview to a more detailed examination of comments within the comment map, the sheer volume of comments can pose challenges in terms of navigation. To mitigate this issue, one possible solution is to employ comment sampling or aggregate similar comments within a single hexagonal grid. Second, it is important to highlight that our approach intentionally omits the visual depiction of intricate connections between comments. This deliberate omission aims to prevent overwhelming visual clutter within the interface. Third, it is crucial to acknowledge that the mechanism prediction model employed in our study is trained using a manually labeled training dataset. However, it is essential to recognize that this training set might be limited in terms of sample size and potential biases, which could potentially impact the generalizability of the model's predictions. \clf{Fourth, the evaluation of the visualizations' effectiveness relies on the feedback obtained from participants involved in the user study. While none of the participants expressed significant concerns, we encourage future work to include quantitative experiments to further assess the efficacy of the visualization design.}


\section{Conclusion}
\par In this research, we aim to enhance the music listening experience using music comments. To achieve this goal, we conducted a need-finding questionnaire and interviews to comprehend the challenges and requirements of music listeners in exploring songs and comments. Subsequently, we integrated two features into an existing music app for music exploration. We evaluated our design through a usage scenario and a user study. Our findings reveal that users demand a more personalized music listening experience, and providing additional comment-related information and well-designed data displays promotes text-based community sharing, communication, and interaction.

\balance
\bibliographystyle{abbrv-doi-hyperref-narrow}

\bibliography{template}

\end{document}